\documentclass[
a4paper,%
12pt,%
twoside%
]{article}

\title{Bremsstrahlung in alpha-decay: angular analysis of
spectra\footnote{
The paper is published in Journ.
Problems of atomic science and technology. 2004, Vol.~5.
\it{Series:} Nuclear Physics Investigations (44), p.~19-21}}

\author{
Sergei~P.~Maydanyuk \thanks{E-mail: maidan@kinr.kiev.ua},
Sergei~V.~Belchikov \thanks{E-mail: sbelchik@kinr.kiev.ua} \\
\small\emph{Institute for Nuclear Research,
National Academy of Sciences of Ukraine,} \\
\small\emph{prosp. Nauki, 47, Kiev-28, 03680, Ukraine}}

\date{}
\begin{document}
\begin{sloppypar}
\maketitle
%---------------------------------------------------------------------------

%---------------------------------------------------------------------------
\vspace{-10mm}
\begin{abstract}
A quantum mechanical method of calculation of bremsstrahlung spectra
in alpha-decay of heavy nuclei with taking into account an angle
between directions of the alpha-particle motion and the photon
emission is presented. Dependence between the bremsstrahlung
spectrum and the angle value is obtained in a simple analytical
form. The method can be used for a comparative analysis of experimental
data, obtained at different angles.
\end{abstract}
%---------------------------------------------------------------------------

%---------------------------------------------------------------------------

% ***************** 03.XX - Квантовая механика ******************************
% 03.40.Kf - Waves and wave propagation: general mathematical aspects;
%
% ************** 11.XX - Теории частиц и полей ******************************
% 12.20.-m Quantum electrodynamics
% 41.20.Jb - Electromagnetic wave propagation; radiowave propagation.
% 41.60.-m Radiation by moving charges
%
% ***************** 20.XX - Ядерная физика **********************************
% 21.10.Tg Lifetimes
% 23.60.+e Alpha decay
% 23.90.+w Other topics in radioactive decay and in-beam spectroscopy
% 24.10.-i Nuclear-reaction models and methods


% ***************************************************************************
% \normalsize
% \vspace{11mm}
\section{Introduction
\label{sec.0}}

Research of bremsstrahlung in $\alpha$-decay of heavy nuclei has
caused an increased interest last years. We note main purposes for
study of this phenomenon:
\begin{itemize}
\vspace{-1mm}

\item
study of properties of the $\alpha$-decay dynamics on the basis of
analysis of the experimental spectra of the bremsstrahlung;
development of a non-stationary model of detailed description of
this process (including a space barrier region), tested on the basis
of the experimental data;

\vspace{-3mm}
\item
investigation of the \emph{subbarrier bremsstrahlung in the
alpha-decay}, i.~e. the photons emission during tunneling of the
alpha-particle through decay barrier;

\vspace{-3mm}
\item
construction of a method of tunneling time determination of the
alpha-particle through the decay barrier (which approximately equals
to nuclear times values $10^{-20}$--$10^{-24}$~sec) on the basis
of analysis of the experimental spectra of the bremsstrahlung.
\end{itemize}

For successful realization of all these researches the theoretical
model is needed, which allows describing the alpha decay with
bremsstrahlung and calculating their main characteristics, and
which is tested by experimental data.

The experiments were fulfilled with such nuclei: $^{210}\mbox{Po}$,
$^{214}\mbox{Po}$, $^{226}\mbox{Ra}$ and $^{244}\mbox{Cm}$. Note,
that in the behavior of the experimental spectra for the nucleus
$^{210}\mbox{Po}$, obtained by Russian-Italian group
\cite{Eremin.2000.PRLTA} and Japanese group \cite{Kasagi.1997.PRLTA}
independently, there is a qualitative difference, which had caused
discussions in some papers. These experiments were fulfilled for the
different values of the angle between the directions of the
alpha-particle motion and the photon emission (which equal to
$90^{\circ}$ and $25^{\circ}$) and, perhaps, by this fact one can
explain the difference between their spectra.
But we note that more accurate analysis can be fulfilled on the
basis of the unified model which allows calculating the
bremsstrahlung spectra with taking into account the different
values of such angle, and this analysis had not done else.

Constructed theoretical models are differed also in their
description of the bremsstrahlung spectra.
Here, the instant accelerated model \cite{D'Arrigo.1994.PHLTA},
developed on the basis of classical electrodynamics and used
such characteristics as a velocity of the alpha-particle during
its leaving outside from the barrier region, gives enough good
description of the experimental bremsstrahlung spectra. But one can
consider these characteristics as additional parameters, which
introduction allows moving the calculated bremsstrahlung spectrum
curve near the experimental data.
Quantum mechanical models, proposed both by T.~Papenbrock and
G.~F.~Bertsch \cite{Papenbrock.1998.PRLTA}, and by E.~V.~Tkalya
\cite{Tkalya.1999.PHRVA}, do not use such additional parameters and
their descriptions of experimental spectra are less satisfactory
(from our point of view).

However, the models, constructed on the basis of quantum
electrodynamics and without semi-classical approach, are the most
effective in study of the alpha-decay dynamics in the space barrier
region and also for detailed study of the subbarrier bremsstrahlung
effect in the alpha-decay.
And the model, in which the method of calculation of the
bremsstrahlung spectra takes into account the angle value, will be
useful for analysis of the experimental spectra, obtained for
different angles (at present, the proposed models
\cite{Kasagi.1997.PRLTA,D'Arrigo.1994.PHLTA,Papenbrock.1998.PRLTA,Tkalya.1999.PHRVA} 
are isotropic).
% ***************************************************************************

% ***************************************************************************
% \normalsize
\section{A short review of quantum mechanical models
\label{sec.1}}

In paper \cite{Maydanyuk.2003.PTP} we proposed the multipolar
quantum mechanical model of the alpha-decay with the bremsstrahlung,
which allows calculating the bremsstrahlung spectra in dependence
on the angle value between directions of the alpha-particle motion and
the photons emission. In accordance with this model, the probability
of the spontaneous photon emission in the alpha-decay is:
\begin{equation}
%\begin{eqnarray}
\begin{array}{lcl}
  \displaystyle\frac{dW}{d\Omega_{\nu}} & = &
        \displaystyle\frac{Z^{2}_{\rm eff}e^{2}k_{f}w_{fi}}{(2\pi)^{4}m}
        \bigl| p(k_{i}, k_{f}) \bigr|^{2}, \\
  k_{i,f} & = & \displaystyle\frac{1}{\hbar}\sqrt{2mE_{i,f}}, \\
  w_{fi} & = & E_{i} - E_{f},
\end{array}
\label{eq.1.1}
% \end{eqnarray}                                          %       (1.1)
\end{equation}
where $p(k_{i}, k_{f})$ has the following form 
\begin{equation}
  p(k_{i}, k_{f}) =
        \sum\limits_{\alpha = 1, 2}
        {\bf e}^{(\alpha)*}
        \int\limits^{+\infty}_{0} dr \int d\Omega
        r^{2} \psi^{*}_{f}({\bf r})
        e^{-i{\bf kr}}
        \displaystyle\frac{\partial}{\partial {\bf r}}
        \psi_{i}({\bf r}).
\label{eq.1.2}
\end{equation}                                          %       (1.1)
Here
$Z_{\rm eff}$ is the effective charge,
$m$ is reduced mass of the composite system (the alpha-particle and
the daughter nucleus),
$E_{i}$ and $E_{f}$ are the total energy of the system in initial 
$i$-state (i.~e. the state of the system before the photon emission) and
final $f$-state (i.~e. the state of the system after the photon emission),
$k_{i}$ and $k_{f}$ are
the wave vector of the system in the initial $i$- and final $f$-states,
$\psi_{i}(\mathbf{r})$ and $\psi_{f}(\mathbf{r})$ are
the wave function of the system in the initial $i$- and final $f$-states,
$\mathbf{e}^{(\alpha)}$ is the unit polarization vector of the emitted
photon,
$\mathbf{k}$ is the wave vector of the photon,
$w_{if} = k = \bigl|\mathbf{k}\bigr|$.
Vector $\mathbf{e}^{(\alpha)}$ is perpendicular to $\mathbf{k}$ in
Coulomb calibration.
We use the unit system when $\hbar = 1$ and $c = 1$.

Main difference between the methods, in which the calculation of the
bremsstrahlung spectrum is based on quantum electrodynamics, consists
in different approaches for calculation of the value $p(k_{i}, k_{f})$
(on our view, the best review of such methods is present in
\cite{Tkalya.1999.PHRVA}).
Here, we use the multipolar expansion of the vector potential of
electromagnetic field of the daughter nucleus (see~\cite{Eisenberg.1973},
p.~57, 51, 49).
Our result is: 
\begin{equation}
\begin{array}{lcl}
  p(k_{i}, k_{f}) & = &
        \sqrt{2\pi} \sum\limits_{l=1} \biggl(\sqrt{2l+1} (-i)^{l}
        [p^{Ml}(k_{i}, k_{f}) - ip^{El}(k_{i}, k_{f})] \biggr), \nonumber \\
  p^{Ml}(k_{i}, k_{f}) & = &
        I_{1}J_{l}(l),                \nonumber                          \\
  p^{El}(k_{i}, k_{f}) & = &
        -\sqrt{\displaystyle\frac{l+1}{2l+1}} I_{2}J_{l}(l-1) +
         \sqrt{\displaystyle\frac{l}{2l+1}}   I_{3}J_{l}(l+1),
\end{array}
\label{eq.1.3}
\end{equation}
where
\begin{equation}
  J_{l}(n) = \int\limits^{+\infty}_{0}
        r^{2}\psi^{*}_{f,l}(r)
        \displaystyle\frac{d\psi_{i}(r)}{dr}
        j_{n}(kr) dr,
\label{eq.1.4}
\end{equation}                                          %       (3.2.12)
\begin{eqnarray}
  I_{1} & = &
        \sum\limits_{\mu=-1}^{1} \mu\hbar_{\mu} \int
        Y_{LM}^{*}({\bf n} ^{f}_{r})
        {\bf T}_{01,0}({\bf n} ^{i}_{r})
        {\bf T}_{ll,\mu}^{*}({\bf n} _{\nu}) d\Omega,\nonumber 
       \\
  I_{2} & = &
        \sum\limits_{\mu=-1}^{1} \mu^{2}\hbar_{\mu} \int
        Y_{LM}^{*}({\bf n} ^{f}_{r})
        {\bf T}_{01,0}({\bf n} ^{i}_{r})
        {\bf T}_{ll-1,\mu}^{*}({\bf n} _{\nu}) d\Omega,
        \nonumber \\
  I_{3} & = &
        \sum\limits_{\mu=-1}^{1} \mu^{2}\hbar_{\mu} \int
        Y_{LM}^{*}({\bf n} ^{f}_{r})
        {\bf T}_{01,0}({\bf n} ^{i}_{r})
        {\bf T}_{ll+1,\mu}^{*}({\bf n}_{\nu}) d\Omega,
\label{eq.1.5}
\end{eqnarray}                                          %       (3.2.13)
and $j_{n}(kr)$ is the spherical Bessel function of order $n$ 
($n$ is a natural number),
$Y_{lm}(\mathbf{n}^{f}_{r})$ are the normalized spherical functions,
$\mathbf{T}_{ll',\mu}(\mathbf{n})$ are the vector spherical harmonics
(see~\cite{Eisenberg.1973}, p.~45).

Calculations of the bremsstrahlung spectrum in the alpha-decay of
the nucleus $^{210}\mbox{Po}$ on the basis of such a model give enough
good description of the experimental data \cite{Eremin.2000.PRLTA},
obtained by Russian-Italian group for the angle $90^{\circ}$.
Evaluation of the spectrum at angle $25^{\circ}$ on the basis
of this models gives a monotonous behaviour (without appearance
of the ``hole'' like experimental data \cite{Kasagi.1997.PRLTA}).

However, there are some problems in the described model, concerned
with the restriction of calculations accuracy of wave functions in
the asymptotic area in the states before and after the photon
emission. As the subintegral expression in the integral
(\ref{eq.1.4}) is the oscillated and slowly damped function, that
calculations convergence of the spectra is limited. As a result,
the numerical calculation of the spectra becomes difficult
sufficiently and in this sense the model \cite{Maydanyuk.2003.PTP}
is not suitable enough.
The convergence of the spectra calculation by the models
\cite{Papenbrock.1998.PRLTA,Tkalya.1999.PHRVA} is higher and these
models are not exposed to necessity of calculations of the wave
functions in the asymptotic area with high precision.
But, probably, they give more approximated calculation of the
bremsstrahlung spectra (see~\cite{Maydanyuk.2003.PTP}).

The second important point in the analysis of quantum mechanical
models is their possibility to calculate the bremsstrahlung
spectra in dependence on the value of the angle between directions
of the alpha-particle motion and the photons emission, and, as result,
a possibility to fulfill a comparative analysis of the experimental
data for the nucleus $^{210}\mbox{Po}$
\cite{Eremin.2000.PRLTA,Kasagi.1997.PRLTA}, obtained for the angles
$90^{\circ}$ and $25^{\circ}$.
Here, the model \cite{Maydanyuk.2003.PTP} allows doing this at the
first time.
In this models the dependence of the bremsstrahlung spectra on the
angle value between the directions of the alpha-particle motion and
the photon emission is in the angular integrals (\ref{eq.1.5}).
However, such angular dependence is not suitable for the speed
angular qualitative analysis of experimental spectra.

Further, we present a new alternative approach for the calculation of
the bremsstrahlung spectra, where such angular dependence is shown
more obviously and more simple.
% ***************************************************************************

% ***************************************************************************
% \normalsize
\section{
A simplified approach for the angular calculations of the spectra 
\label{sec.2}}

Let's consider $\bf k$ and $\bf r$.
Vector $\bf k$ is the photon impulse and points to a direction of
the photon motion.

Vector $\bf r$ is the radius-vector, which points to a space position
of the alpha-particle relatively to the center of mass of the daughter
nucleus. We suppose, that as the mass of the daughter nucleus is
sufficiently more then the alpha-particle mass, then the direction
of the radius-vector of the alpha-particle position coincides with
the direction of alpha-particle velocity.
Then the angle between the vectors $\bf k$ and $\bf r$ is the angle
between the directions of the alpha-particle motion and the photon
emission.
One can write:
\begin{equation}
  \exp{(-i{\bf kr})} = \exp{(-ikr \cos{\beta})},
\label{eq.2.1}
\end{equation}                                          %       (2.1)
where
$k = |{\bf k}|$,
$r = |{\bf r}|$,
$\beta$ is the angle between the direction of the alpha-particle
motion $\mathbf{k} / k$ and the direction of the propagation of
the emitted photon $\mathbf{r} / r$.

Find the expression for the bremsstrahlung spectrum.
Let's write polarization vectors $\mathbf{e}^{\alpha}$ in terms of
circular polarization vectors $\mathbf{\xi}$ with opposite
directions of rotation (see~\cite{Eisenberg.1973}, p.~42):
\begin{equation}
\begin{array}{ll}
  {\bf \xi}_{-1} = \displaystyle\frac{1}{\sqrt{2}}
                   (\mathbf{e}^{1} - i\mathbf{e}^{2}), &
  {\bf \xi}_{+1} = -\displaystyle\frac{1}{\sqrt{2}}
                   (\mathbf{e}^{1} + i\mathbf{e}^{2}).
\end{array}
\label{eq.2.2}
\end{equation}                                          %       (2.2)
We obtain:
\begin{equation}
  p(k_{i}, k_{f}) =
        \sum\limits_{\mu = -1, 1}
        h_{\mu}\mathbf{\xi}^{*}_{\mu}
        \int\limits^{+\infty}_{0} dr \int d\Omega
        r^{2} \psi^{*}_{f}(\mathbf{r})
        e^{-i\mathbf{kr}}
        \displaystyle\frac{\partial}{\partial \mathbf{r}}
        \psi_{i}(\mathbf{r}),
\label{eq.2.3}
\end{equation}                                          %       (2.3)
where
\begin{equation}
\begin{array}{ll}
  h_{-1} = \displaystyle\frac{1}{\sqrt{2}} (1 - i), &
  h_{1}  = \displaystyle\frac{1}{\sqrt{2}} (-1 - i).
\end{array}
\label{eq.2.4}
\end{equation}                                          %       (2.4)

We use the following property (see~\cite{Eisenberg.1973}, p.~44-46;
\cite{Maydanyuk.2003.PTP}):
%
% \begin{eqnarray}
\begin{equation}
\begin{array}{l}
  \displaystyle\frac{\partial}{\partial {\bf r}}
  \psi_{i}({\bf r}) =
        -\displaystyle\frac{d\psi_{i}(r)}{dr}
        {\bf T}_{01,0}({\bf n} ^{i}_{r}),       \\
  {\bf T}_{01,0}({\bf n} ^{i}_{r}) =
        \sum\limits^{1}_{\mu = -1} (110|-\mu\mu 0)
        Y_{1,-\mu}({\bf n} ^{i}_{r}) {\bf \xi}_{\mu}, \\
  (110|1, -1, 0) = (110|-1, 1, 0) = \sqrt{\displaystyle\frac{1}{3}},
\end{array}
\label{eq.2.5}
\end{equation}
% \end{eqnarray}                                          %       (2.5)
% 
where
$(110 | -\mu\mu 0)$ are the Clebsch-Gordan coefficients.
Taking into account (\ref{eq.2.5}), (\ref{eq.2.1}) and the orthogonal
property of the polarization vectors $\mathbf{\xi}_{-1}$ and
$\mathbf{\xi}_{1}$, we find:
\begin{equation}
  p(k_{i}, k_{f}) = -
        \sqrt{\displaystyle\frac{1}{3}}
        \sum\limits_{\mu = -1, 1}
        h_{\mu}
        \int\limits^{+\infty}_{0} dr
        r^{2} \psi^{*}_{f}(r)
        \displaystyle\frac{\partial \psi_{i}(r)}{\partial r}
        \int d\Omega
        Y_{l'm'}^{*}(\mathbf{n}^{f}_{r})
        Y_{1,-\mu}(\mathbf{n}^{i}_{r})
        e^{-ikr\cos{\beta}}
\label{eq.2.6}
\end{equation}                                          %       (2.6)

Further, we suppose that the process of the photon creation does
not change the direction of the alpha-particle motion, i.~e.:
\begin{equation}
   \mathbf{n}^{i}_{r} = \mathbf{n}^{f}_{r}.
\label{eq.2.7}
\end{equation}
Taking into account this approximation, and also the orthogonal
property of the functions $Y_{lm}(\mathbf{n}_{r})$, we obtain the
following expression for $p(k_{i}, k_{f})$:
\begin{equation}
  p(k_{i}, k_{f}) =
        \sqrt{\displaystyle\frac{1}{3}}
        \sum\limits_{\mu = -1, 1}
        h_{\mu}
        \int\limits^{+\infty}_{0}
        r^{2} \psi^{*}_{f}(r, l=1, m=-\mu)
        \displaystyle\frac{\partial \psi_{i}(r, l=m=0)}
        {\partial r}
        e^{-ikr \cos{\beta}}
        dr.
\label{eq.2.8}
\end{equation}                                          %       (3.2.1)

Note, that we have the quantum numbers $l=1$, $m=\mu$ for final
state and $l=m=0$ for initial state. Such determination follows
from the supposition (\ref{eq.2.7}). From here we can also
conclude, that in multipolar approach of quantum mechanical
calculation of the bremsstrahlung spectra the E1 multipole gives
the most important contribution into the total spectrum (such idea
of estimation of multipole contributions into a total
bremsstrahlung spectrum is obtained at the first time).

In such a form of $p(k_{i}, k_{f})$ the dependence of the
bremsstrahlung spectrum on the angle is more obvious.
In new approach one can calculate the bremsstrahlung spectrum on
the basis of (\ref{eq.1.1}), where one can use (\ref{eq.2.8}) for
$p(k_{i}, k_{f})$ value.
Note that the model \cite{Maydanyuk.2003.PTP} allows calculating
the bremsstrahlung spectra only with taking into consideration of
the selected electrical and magnetic multiples, whereas the method,
proposed in this paper, gives the calculation of the bremsstrahlung
spectrum as a whole.
% ***************************************************************************

% ***************************************************************************
% \normalsize
% \section{
% Результаты
% \label{sec.3}}
% ***************************************************************************

% ***************************************************************************
% \normalsize
\section{Conclusions
\label{sec.4}}

We propose the new approach for calculation of the
bremsstrahlung spectra in alpha-decay, where the angle between
the directions of the alpha-particle motion and the photon emission
is taken into account, and the dependence of the bremsstrahlung
spectra on the values of such angle is more obvious and simple
then in the model \cite{Maydanyuk.2003.PTP}.
This method can be useful for a comparative analysis of the
experimental data \cite{Eremin.2000.PRLTA,Kasagi.1997.PRLTA}, obtained
for different angles.
% ***************************************************************************

% ***************************************************************************
% \normalsize
%\section*{Благодарности
\section*{Acknowledgements
\label{sec.5}}

The authors are grateful to Prof.~V.~S.~Olkhovsky for setting of
the problem of the bremsstrahlung in the alpha-decay and for
productive discussions concerning of study of this phenomena
and calculations of tunneling times.
% ***************************************************************************

% ***************************************************************************
\bibliographystyle{h-physrev4}

\end{sloppypar}
\end{document}